\documentclass[aps,prl,10pt,twocolumn,twoside,showkeys]{revtex4-2}

\usepackage{amsmath,amsfonts,amssymb,amsthm,mathtools}

\usepackage[english]{babel}
\usepackage[utf8]{inputenc}
\usepackage[T1]{fontenc}

\usepackage{latexsym}
\usepackage[pdftex]{graphicx}
\usepackage{epstopdf}

\usepackage{subfigure}
\usepackage{enumitem}

\usepackage{hyperref}
\usepackage[all]{hypcap}

\hypersetup{
	pdftitle={},
	pdfauthor={},
	colorlinks=true,
	linkcolor=black,
	citecolor=blue,
	filecolor=black,
	urlcolor=blue,
	bookmarksopen,
	bookmarksopenlevel=1,
}

\newcommand{\cX}{\mathcal{X}}

\newcommand{\ee}{\mathrm{e}}
\newcommand{\ii}{\mathrm{i}}
\newcommand{\dd}{\mathrm{d}}

\def\gl{\mathfrak{gl}}
\def\sl{\mathfrak{sl}}

\def\sgn{\operatorname{sgn}}

\def\ad{\operatorname{ad}}

\newcommand{\wick}[1]{\left. :\! \hspace{-0.5pt} #1 \hspace{-0.5pt} \!: \right.}

\newcommand{\anyonWick}[2]{N_{#2}[ #1 ] }

\theoremstyle{definition}

\theoremstyle{remark}

\begin{document}


\title{%
Exact Dirac-Bogoliubov-de Gennes Dynamics for Inhomogeneous Quantum Liquids%
}%

\author{Per Moosavi}
\email{pmoosavi@phys.ethz.ch}
\affiliation{%
Institute for Theoretical Physics, ETH Zurich,
Wolfgang-Pauli-Strasse 27, 8093 Z{\"u}rich, Switzerland%
}%

\date{%
September 6, 2023%
}%

\begin{abstract}
We study inhomogeneous 1+1-dimensional quantum many-body systems described by Tomonaga-Luttinger-liquid theory with general propagation velocity and Luttinger parameter varying smoothly in space, equivalent to an inhomogeneous compactification radius for free boson conformal field theory. This model appears prominently in low-energy descriptions, including for trapped ultracold atoms, while here we present an application to quantum Hall edges with inhomogeneous interactions. The dynamics is shown to be governed by a pair of coupled continuity equations identical to inhomogeneous Dirac-Bogoliubov-de Gennes equations with a local gap and solved by analytical means. We obtain their exact Green's functions and scattering matrix using a Magnus expansion, which generalize previous results for conformal interfaces and quantum wires coupled to leads. Our results explicitly describe the late-time evolution following quantum quenches, including inhomogeneous interaction quenches, and Andreev reflections between coupled quantum Hall edges, revealing a remarkably universal dependence on details at stationarity or at late times out of equilibrium.
\end{abstract}


\maketitle


\setlength{\abovedisplayskip}{7pt}
\setlength{\belowdisplayskip}{7pt}

\emph{Introduction.}---%
\csname phantomsection\endcsname%
\addcontentsline{toc}{section}{Introduction}%
%
Tomonaga-Luttinger liquids (TLLs) \cite{Tomonaga:1950, Luttinger:1963, MattisLieb:1965, Haldane:1981, Haldane:1981b} is a prominent class of gapless quantum many-body systems whose low-energy physics is described by the conformal field theory (CFT) of 1+1-dimensional compactified free bosons.
An important generalization is the inhomogeneous theory where the propagation velocity $v(x)$ and the Luttinger parameter $K(x)$ are positive functions of position $x$.
This was studied for quantum wires connected to leads \cite{MaSt:1995, SaSc:1995, Pono:1995, GGM:2010}, as effective descriptions of trapped ultracold atoms in equilibrium \cite{Stringari:1996, HoMa:1999, MenottiStringari:2002, Ghosh:2004, PetrovEtAl:2004, CitroEtAl:2008}, and recently in nonequilibrium contexts \cite{DZT:2008, CCGOR:2011, GeigerEtAl:2014, DoraPollmann:2015, ADSV:2016, BrunDubail:2018igff, Bastianello:2020ill, RCDD:2020, GMS:2022, RCGF:2022, TajikEtAl:2023}.
How to handle general $v(x)$ is known \cite{DSVC:2017, DSC:2017, GLM:2018, LaMo2:2019, Moo:2021iCFT}, but obtaining solutions for general $K(x)$ is an outstanding problem, in or out of equilibrium.

In this Letter, we approach this problem by showing that the dynamics is governed by two coupled partial differential equations (PDEs) that we solve analytically in full generality, revealing remarkably universal late-time evolution following quantum quenches and the presence of Andreev reflections.
The PDEs are identified as inhomogeneous Dirac-Bogoliubov-de~Gennes (DBdG) equations with an effective local gap $\Delta(x) \equiv v(x) \Lambda(x)$ where
\begin{equation}
\label{Lambda_x}
\Lambda(x)
\equiv
\partial_x \log \sqrt{K(x)}.
\end{equation}
Such equations are well known in superconductivity (but different as our gap has no self-consistency criterion), describing Andreev reflections at superconductor-normal-metal interfaces \cite{Andreev:1964}.
Moreover, they were used to study, e.g., graphene \cite{Been1:2006, Been2:2008}, certain junctions \cite{MaslovEtAl:1996, ChHa:1998, TiBe:2006, TOD:2008}, and fractional quantum Hall (FQH) systems \cite{ReGr:2000}, also with an inhomogeneous velocity \cite{Peres:2009, PPR:2017}.
Studying inhomogeneous TLLs using PDEs is not new \cite{MaSt:1995, SaSc:1995, Pono:1995, GGM:2010}, but their essential form and significance were not recognized before, and, to our knowledge, no one gave the full analytical solution.

Besides its importance for condensed-matter applications, where $K(x)$ encodes interactions in an underlying quantum many-body system, the problem is interesting also in high-energy theory, where $K(x)$ appears as an inhomogeneous compactification radius $R(x) = \sqrt{2\alpha' K(x)}$ ($\alpha'$ has dimension length squared) \cite{Polchinski:1998vol1}.
For stepwise changes in $R(x)$, this was studied using interface operators in boundary CFT \cite{BachasEtAl:2002, BachasBrunner:2008} and analogous operators for stepwise changes in time \cite{DLMT:2023}.
However, general $R(x)$ were not considered.

\begin{figure}[!htbp]

\centering

\includegraphics[scale=1, trim=0 0 0 0, clip=true]{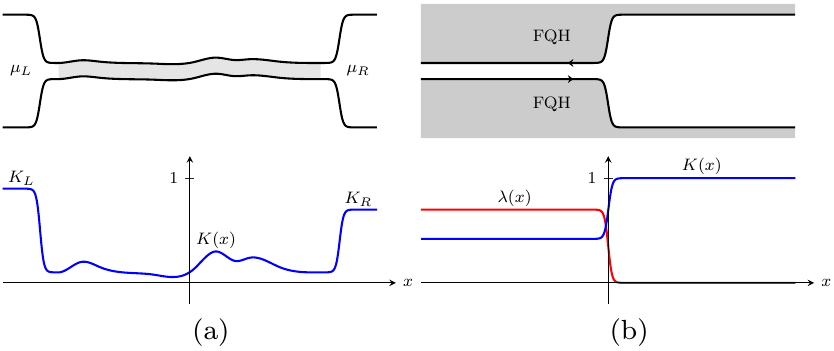}

\vspace{-4mm}

\caption{%
Illustrations of (a) a quantum wire coupled to leads with chemical potentials $\mu_{L,R}$, and (b) coupled FQH edges modeled as CFTs of counterpropagating anyons with inhomogeneous density-density interaction $\lambda(x) = \frac{1 - K(x)^2}{1 + K(x)^2}$.%
}%
\label{Fig:Illustrations}

\end{figure}

Our main results are the exact Green's functions and scattering matrix for the governing DBdG equations, expressible using a Magnus expansion in a natural interaction picture.
These are fully explicit at stationarity or at late times out of equilibrium and describe the full time evolution perturbatively in $\Lambda(x)$.
They also explain, from a PDE perspective, the predicted breaking of the Huygens-Fresnel principle \cite{GMS:2022} and Andreev reflections \cite{SaSc:1995, MaslovGoldbart:1998, DZT:2008} in inhomogeneous TLLs.
Physical applications include quasi-1+1-dimensional ultracold gases, gapless quantum $XXZ$ spin chains with smoothly varying couplings \cite{DoraPollmann:2015, DSC:2017, Moo:2021iCFT, GMS:2022}, and quantum wires, see Fig.~\ref{Fig:Illustrations}(a).
As a new application, we present a toy model for Andreev reflections between coupled FQH edges described by anyonic CFTs with inhomogeneous density-density interactions, which we map to an inhomogeneous TLL, see Fig.~\ref{Fig:Illustrations}(b).
This is motivated by recent experimental observation of Andreev reflections between FQH edges coupled via a superconductor \cite{GulEtAl:2020}, cf.\ also \cite{SCF:1998, NFLL:1999, HJASRMM:2021, SchillerEtAl:2023}.
We expect our results to have wide importance and
applicability, not only for TLLs but whenever \mbox{(Dirac-)}Bogoliubov-de~Gennes-type equations appear, including superconductor interfaces in higher dimensions and continuum descriptions \cite{TLM:1980} of the Su-Schrieffer-Heeger model \cite{SSH:1979}.

\emph{Inhomogeneous TLLs.}---%
\csname phantomsection\endcsname%
\addcontentsline{toc}{section}{Inhomogeneous TLLs}%
%
Inhomogeneous TLL theory on the circle $S^1 = [-\frac{L}{2}, \frac{L}{2}]$ can be formulated in terms of a compactified bosonic field $\varphi(x)$ (modulo $2\pi$) with conjugate $\Pi(x)$ for $x \in S^1$ satisfying $[\partial_x \varphi(x), \Pi(y)] = \ii \delta'(x-y)$ and periodic boundary conditions, where $\delta(x)$ is the $L$-periodic delta function.
The inhomogeneities are modeled by periodic positive functions $v(x)$ and $K(x)$ that, for completeness, we assume are smooth.
The Hamiltonian is (setting $\hbar = 1$) \cite{Note:Generic_iTLL_H}
\begin{equation}
\label{H_iTLL}
H
\equiv
\int_{S^1} \frac{\dd x}{2\pi} v(x)
  \! \wick{
    \biggl(
      \frac{[\pi \Pi(x)]^2}{K(x)}
      + K(x)[\partial_x \varphi(x)]^2
    \biggr)
  } \! ,
\end{equation}
up to subtracting the Casimir contribution $\int_{S^1} \dd x\, \frac{\pi v(x)}{6L^2}$ (which is suppressed for simplicity).
Here, $\wick{ \cdots }$ denotes Wick ordering, which can be defined in analogy with the homogeneous case by expressing $H$ in terms of oscillator modes identified by expanding the fields in appropriate eigenfunctions obtained from an associated Sturm-Liouville problem \cite{Stringari:1996, HoMa:1999, MenottiStringari:2002, Ghosh:2004, PetrovEtAl:2004, CitroEtAl:2008, GMS:2022}.
(Although a viable option, we will not follow that approach here.)
As this ordering amounts to a shift by a constant in \eqref{H_iTLL}, it will be of no consequence for the results presented here.

To study the evolution in time $t$, we instead write
\begin{equation}
H
= \int_{S^1} \dd x\,
  \pi v(x) \!
  \wick{
    \Bigl(
      \tilde{\rho}_{+}(x)^2
      + \tilde{\rho}_{-}(x)^2
    \Bigr)
  }
\end{equation}
using a $K(x)$-dependent partitioning into right- ($+$) and left- ($-$) moving plasmon  densities
\begin{equation}
\label{trho_pm}
\tilde{\rho}_{\pm}(x)
\equiv
\frac{1}{2\pi}
\Biggl[ \frac{1}{\sqrt{K(x)}} \pi \Pi(x) \mp \sqrt{K(x)} \partial_x \varphi(x) \Biggr],
\end{equation}
which can be shown to satisfy
\begin{subequations}
\label{trho_comm_rels}
\begin{align}
[\tilde{\rho}_{\pm}(x), \tilde{\rho}_{\pm}(y)]
& = \mp \frac{\ii}{2\pi} \delta'(x-y),
    \label{trho_comm_rels_a} \\
[\tilde{\rho}_{+}(x), \tilde{\rho}_{-}(y)]
& = \frac{\ii}{2\pi} \Lambda(x) \delta(x-y),
    \label{trho_comm_rels_b}
\end{align}
\end{subequations}
the latter featuring the new coupling $\Lambda(x)$ defined in \eqref{Lambda_x}.
The densities $\tilde{\rho}_{\pm}(x)$ together with the associated currents
\begin{equation}
\label{tj_pm}
\tilde{j}_{\pm}(x)
\equiv
\pm v(x) \tilde{\rho}_{\pm}(x)
\end{equation}
will be shown to satisfy coupled continuity equations
\begin{equation}
\label{cont_eqs_tilde}
\partial_t \tilde{\rho}_{\pm} + \partial_x \tilde{j}_{\pm}
= \pm v(x)\Lambda(x) \tilde{\rho}_{\mp}.
\end{equation}
After reformulating \eqref{cont_eqs_tilde} into inhomogeneous DBdG equations, our approach will be to solve them directly.

Before continuing, we find it worth noting that
$J_{n} \equiv \int_{S^1} \dd x\, J_{+}(x) \ee^{-2\pi \ii n x/L}$
and
$\bar{J}_{n} \equiv \int_{S^1} \dd x\, J_{-}(x) \ee^{2\pi \ii n x/L}$
for
$J_{\pm}(x) \equiv \sqrt{K(x)} \tilde{\rho}_{\pm}(x)$
obey natural generalizations of algebraic relations for standard TLLs (compactified free bosons).
Indeed, from \eqref{trho_comm_rels}, one obtains new coupled $\mathrm{U}(1)$ current algebras,
\begin{gather}
[J_{n}, J_{m}]
= \frac{n-m}{2} K_{n+m},
\quad
[\bar{J}_{n}, \bar{J}_{m}]
= \frac{n-m}{2} K_{-n-m}, \nonumber \\
[J_{n}, \bar{J}_{m}]
= \frac{m-n}{2} K_{n-m},
\label{cdu1_current_alg}
\end{gather}
where $K_{n} \equiv L^{-1} \int_{S^1} \dd x\, K(x) \ee^{-2\pi \ii n x/L}$ is dimensionless and couples the otherwise commuting algebras for right and left movers:
If $K(x) = K$, then $K_{n} = K \delta_{n,0}$, and the algebras decouple.

\emph{Charge transport and DBdG equations.}---%
\csname phantomsection\endcsname%
\addcontentsline{toc}{section}{Charge transport and DBdG equations}%
%
Using the Heisenberg equation, one finds that the model in \eqref{H_iTLL} has two conserved particle currents with
\begin{subequations}
\label{rho_j_rho5_j5}
\begin{align}
\rho(x)
& = \Pi(x),
& \jmath(x)
& = v(x)K(x) \rho_{5}(x),
    \label{rho_j_rho5_j5_a} \\
\rho_{5}(x)
& = - \partial_{x} \varphi(x)/\pi,
& \jmath_{5}(x)
& = v(x) K(x)^{-1} \rho(x),
    \label{rho_j_rho5_j5_b}
\end{align}
\end{subequations}
satisfying $\partial_{t} \rho + \partial_{x} \jmath = 0$, $\partial_{t} \rho_{5} + \partial_{x} \jmath_{5} = 0$, and
\begin{subequations}
\label{j_j5_PDEs_2}
\begin{align}
\partial_{t} \jmath + v(x)K(x) \partial_{x} [v(x) K(x)^{-1} \rho]
& = 0, \\
\partial_{t} \jmath_{5} + v(x)K(x)^{-1} \partial_{x} [v(x)K(x) \rho_{5}]
& = 0.
\end{align}
\end{subequations}
These can be recast into the coupled continuity equations in \eqref{cont_eqs_tilde} for $\tilde{\rho}_{\pm}$ in \eqref{trho_pm} and $\tilde{j}_{\pm}$ in \eqref{tj_pm}, where the latter allow one to write the particle density and its associated current as
$\rho = \sqrt{K(x)} \bigl( \tilde{\rho}_{+} + \tilde{\rho}_{-} \bigr)$
and
$\jmath = \sqrt{K(x)} \bigl( \tilde{j}_{+} + \tilde{j}_{-} \bigr)$.
In turn, it is straightforward to show that these coupled equations are equivalent to the inhomogeneous DBdG equations
\begin{equation}
\label{iDBdG_eqs}
\begin{pmatrix}
  v(x) \partial_x + \partial_t & \Delta(x) \\
  \Delta(x) & v(x) \partial_x - \partial_t
\end{pmatrix}
\begin{pmatrix}
  \tilde{j}_{+} \\
  \tilde{j}_{-}
\end{pmatrix}
= \begin{pmatrix}
    0 \\
    0
  \end{pmatrix}
\end{equation}
with the local gap $\Delta(x) = v(x) \Lambda(x)$ given by \eqref{Lambda_x}.
One of our main messages is that the dynamics in any inhomogeneous TLL is governed by these equations.

\emph{Solution in the infinite volume.}---%
\csname phantomsection\endcsname%
\addcontentsline{toc}{section}{Solution in the infinite volume}%
%
The rest of this Letter is dedicated to solving \eqref{iDBdG_eqs} and analyzing the solutions.
To this end, we consider expectations $\langle \cdot \rangle$ with respect to an arbitrary state, take the infinite-volume limit $L \to \infty$ (thus avoiding questions about boundary conditions), and write the equations in frequency space:
\begin{equation}
\label{iDBdG_eqs_omega}
\bigl[ \partial_{x} - \ii \mathsf{P}_{\omega}(x) \bigr]
\begin{pmatrix}
  \langle \hat{j}_{+}(x, \omega) \rangle \\
  \langle \hat{j}_{-}(x, \omega) \rangle
\end{pmatrix}
= \frac{1}{v(x)} \sigma_{3}
  \begin{pmatrix}
    \langle \tilde{j}_{+}(x, 0) \rangle \\
    \langle \tilde{j}_{-}(x, 0) \rangle
  \end{pmatrix}
\end{equation}
for $x \in \mathbb{R}$, where we have introduced the $2 \times 2$-matrix
\begin{equation}
\label{P_omega_x}
\mathsf{P}_{\omega}(x)
\equiv
  \frac{\omega}{v(x)} \sigma_{3} + \ii \Lambda(x) \sigma_{1}.
\end{equation}
(Here and in what follows, $\sigma_{0,1,2,3}$ denote Pauli matrices.)
This matrix lies in the Lie algebra $\sl(2, \mathbb{C})$ and resembles a parity-time-(anti)symmetric \cite{Bender:2007} non-Hermitian two-level system with free part $\frac{\omega}{v(x)} \sigma_{3}$ and interaction $\ii \Lambda(x) \sigma_{1}$.
To obtain \eqref{iDBdG_eqs_omega}, we assumed a system prepared in an initial state for $t < 0$ and allowed to evolve for $t > 0$ with initial data given by the expectations $\langle \tilde{j}_{\pm}(x, 0) \rangle$ at $t = 0$.
The corresponding conventions for the Fourier transforms are $\hat{j}_{\pm}(x, \omega) = \int_{0}^{\infty} \dd t\, \tilde{j}_{\pm}(x, t) \ee^{\ii \omega t}$.
Solving \eqref{iDBdG_eqs_omega} is nontrivial since $\mathsf{P}_{\omega}(x) \mathsf{P}_{\omega}(y) \neq \mathsf{P}_{\omega}(y) \mathsf{P}_{\omega}(x)$ in general.
As we will see, this requires spatial ordering, analogous to the familiar ordering for time-dependent Hamiltonians.

\emph{Green's functions.}---%
\csname phantomsection\endcsname%
\addcontentsline{toc}{section}{Green's functions}%
%
Suppose $\langle \tilde{j}_{\pm}(x, 0) \rangle$ have compact support and $\langle \tilde{j}_{\pm}(x, t) \rangle \to 0$ as $|x| \to \infty$.
Then the solutions to the DBdG equations are (see the Supplemental Material \cite{SM} for details)
\begin{equation}
\label{j_pm_solutions}
\begin{pmatrix}
  \langle \tilde{j}_{+}(x, t) \rangle \\
  \langle \tilde{j}_{-}(x, t) \rangle
\end{pmatrix}
= \int_{\mathbb{R}} \dd y\,
  G(x, y; t)
  \frac{1}{v(y)}
  \begin{pmatrix}
	  \langle \tilde{j}_{+}(y, 0) \rangle \\
	  \langle \tilde{j}_{-}(y, 0) \rangle
  \end{pmatrix}
\end{equation}
using
$G(x, y; t)
= \int_{\mathbb{R}} \frac{\dd \omega}{2\pi}\, \hat{G}(x, y; \omega) \ee^{-\ii \omega t}$
with
$\hat{G}(x, y; \omega)
= \hat{G}_{+}(x, y; \omega) \frac{\sigma_{0} + \sigma_{3}}{2}
+ \hat{G}_{-}(x, y; \omega) \frac{\sigma_{0} - \sigma_{3}}{2}$
given by ordered exponentials
\begin{equation}
\label{G_pm_xy_omega}
\hat{G}_{\pm}(x, y; \omega)
= \pm \theta(\pm[x-y])
  \overset{\substack{\leftarrow \vspace{-1mm} \\ \rightarrow}}{\cX}
  \ee^{\ii \int_{y}^{x} \dd s\, \mathsf{P}_{\omega}(s)} \sigma_{3}.
\end{equation}
Here, $\overset{\leftarrow}{\cX}$ ($\overset{\rightarrow}{\cX}$) denotes spatial ordering where positions decrease (increase) from left to right.
The interpretation of
$G_{+(-)}(x, y; t)
= \int_{\mathbb{R}} \frac{\dd \omega}{2\pi}\, \hat{G}_{+(-)}(x, y; \omega) \ee^{-\ii\omega t}$
is as a spatially retarded (advanced) Green's function \cite{Note:Propagation_classification}.
Multiplication by $\frac{\sigma_{0} \pm \sigma_{3}}{2}$ in $\hat{G}(x, y; \omega)$ projects the solution to be causal, i.e., propagating forward in time, without which it contains both forward and backward propagation.

If $K(x)$ is constant, \eqref{G_pm_xy_omega} simplifies to
$\hat{G}_{\pm}^{0}(x, y; \omega)
= \pm \theta(\pm[x-y]) \ee^{\ii \omega \tau_{x, y} \sigma_{3}} \sigma_{3}$
with $\tau_{x, y} = \int_{y}^{x} \dd s\, \frac{1}{v(s)}$.
The corresponding $G^{0}(x, y; t)$ readily reproduces the Green's functions in \cite{LaMo2:2019} (before disorder averaging), as expected since the DBdG equations in \eqref{iDBdG_eqs} decouple.

While the Green's functions are nontrivial to compute due to the spatial ordering, there are perturbative and sometimes even exact evaluation schemes \cite{BCOR:2009}.
Indeed, since $\mathsf{P}_{\omega}(x) \in \sl(2, \mathbb{C})$, generated by $\sigma_{3}$ and $\frac{\sigma_{1} \pm \ii \sigma_{2}}{2}$, the exact
$\overset{\substack{\leftarrow \vspace{-1mm} \\ \rightarrow}}{\cX}
	\ee^{\ii \int_{y}^{x} \dd s\, \mathsf{P}_{\omega}(s)}$
can be represented as products of exponentials of these generators with coefficients obtained from solving a Riccati equation by quadrature \cite{WeiNorman:1963, WeiNorman:1964}.
While solvable in principle, we instead use an interaction-picture [cf.\ \eqref{P_omega_x}] Magnus expansion, yielding formulas perturbative in $\Lambda(x)$ \cite{BCOR:2009, Note:Time_rev_inv}:
\begin{equation}
\label{int_pic_Magnus_exp}
\overset{\substack{\leftarrow \vspace{-1mm} \\ \rightarrow}}{\cX}
\ee^{\ii \int_{y}^{x} \dd s\, \mathsf{P}_{\omega}(s)}
= \exp
  \Biggl[ \,
    \sum_{n = 1}^{\infty}
    \Omega_{\omega}^{(n)}(x, y; x)
  \Biggr]
  \ee^{\ii \omega \tau_{x, y} \sigma_{3}}
\end{equation}
with
\begin{subequations}
\label{int_pic_Magnus_coeffs}
\begin{align}
\Omega_{\omega}^{(1)}(x, y; a)
& = \int_{y}^{x} \dd s\, \ii \mathsf{P}_{\omega}^{1}(s; a), \\
\Omega_{\omega}^{(n)}(x, y; a)
& = \sum_{k=1}^{n-1} \frac{B_{k}}{k!}
    \sum_{\substack{m_1 \geq 1, \ldots, m_k \geq 1 \\ m_1 + \ldots + m_k = n-1}}
    \int_{y}^{x} \dd s\, \nonumber \\
& \hspace{-6mm} \times
    \ad_{\Omega_{\omega}^{(m_1)}(s, y; a)} \!
    \ldots
    \ad_{\Omega_{\omega}^{(m_k)}(s, y; a)} \!
    \ii \mathsf{P}_{\omega}^{1}(s; a)
\end{align}
\end{subequations}
for $n \geq 2$ using $\mathsf{P}_{\omega}^{1}(s; a) \equiv \ii \Lambda(s) \begin{pmatrix} 0 & \ee^{-2\ii \omega \tau_{s, a}} \\ \ee^{2\ii \omega \tau_{s, a}} & 0 \end{pmatrix}$ ($a \in \mathbb{R}$) and the Bernoulli numbers $B_k$ (with $B_{1} = -{1}/{2}$).
We stress that the zero-frequency contribution is straightforward to compute since $\mathsf{P}_{0}(x) = \ii \Lambda(x) \sigma_{1}$
for different arguments commute:
The only nonzero contribution in \eqref{int_pic_Magnus_exp} is
$\exp \bigl[ - \int_{y}^{x} \dd s\, \Lambda(s) \sigma_{1} \bigr]
\equiv \mathsf{T}(x, y)$
with
\begin{equation}
\label{T-matrix_xy}
\mathsf{T}(x, y)
= \begin{pmatrix}
    \frac{\sqrt{\frac{K(y)}{K(x)}}+\sqrt{\frac{K(x)}{K(y)}}}{2}
    & \frac{\sqrt{\frac{K(y)}{K(x)}}-\sqrt{\frac{K(x)}{K(y)}}}{2} \\
    \frac{\sqrt{\frac{K(y)}{K(x)}}-\sqrt{\frac{K(x)}{K(y)}}}{2}
    & \frac{\sqrt{\frac{K(y)}{K(x)}}+\sqrt{\frac{K(x)}{K(y)}}}{2}
  \end{pmatrix},
\end{equation}
identified below as the zero-frequency transfer matrix.

To analyze \eqref{j_pm_solutions} at late times, it is instructive to compare $G(x, y; t)$ with $G^{0}(x, y; t)$ for constant $K(x)$:
The difference $G(x, y; \lambda t) - \mathsf{T}(x, y) G^{0}(x, y; \lambda t)$ formally vanishes as $o(\lambda^{-1})$ for $\lambda \gg 1$, which can be shown using \eqref{j_pm_solutions}--\eqref{int_pic_Magnus_coeffs} and rescaling $\omega$ in the inverse Fourier transform by $1/\lambda$ (see the Supplemental Material \cite{SM} for details).
This implies that the leading large-$t$ contribution to $G(x, y; t)$ is expressible as $\mathsf{T}(x, y) G^{0}(x, y; t)$, which is explicitly computable and has a remarkably simple dependence on $K(x)$ via \eqref{T-matrix_xy}.
An important example is the current $\jmath$ in \eqref{rho_j_rho5_j5_a}, for which one obtains
\begin{multline}
\label{j_xt_late_times}
\langle \jmath(x, t) \rangle
= \int_{\mathbb{R}} \dd y\,
  \frac{\delta(\tau_{x, y} - t) - \delta(\tau_{x, y} + t)}{2}
  \langle \rho(y, 0) \rangle \\
  + \int_{\mathbb{R}} \dd y\,
    \frac{\delta(\tau_{x, y} - t) + \delta(\tau_{x, y} + t)}{2v(y)}
    \langle \jmath(y, 0) \rangle
  + o(t^{-1})
\end{multline}
for all $K(x)$.
The density $\langle \rho(x, t) \rangle$ has a similar formula, obtained by inserting $\frac{K(x)}{K(y)v(x)}$ inside the integrals and exchanging $\langle \rho(y, 0) \rangle$ and $\frac{\langle \jmath(y, 0) \rangle}{v(y)}$.
As we will discuss, this exemplifies the universality of the late-time dynamics following a quantum quench, e.g., changing an external potential or modulating an interaction encoded by $K(x)$, underscoring the usefulness of our solution in \eqref{j_pm_solutions}--\eqref{int_pic_Magnus_coeffs}.

Complementary to \eqref{int_pic_Magnus_exp}--\eqref{int_pic_Magnus_coeffs}, a corresponding Magnus expansion in $\omega$ can be obtained by swapping the identifications as free and interaction terms in \eqref{P_omega_x}, verifying the above late-time dynamics.

\emph{Transfer and scattering matrices.}---%
\csname phantomsection\endcsname%
\addcontentsline{toc}{section}{Transfer and scattering matrices}%
%
Consider a scenario where $\langle \tilde{j}_{\pm}(x, 0) \rangle = 0$ for a subsystem on the finite interval $[x_1, x_2]$ and currents instead incident on the boundaries at $x_{1,2}$.
The transfer matrix $\mathsf{T}(\omega)$ corresponding to \eqref{iDBdG_eqs_omega} that connects
$( \hat{j}_{+}(x_1, \omega), \hat{j}_{-}(x_1, \omega) )^{T}$
and
$( \hat{j}_{+}(x_2, \omega), \hat{j}_{-}(x_2, \omega) )^{T}$
is (see the Supplemental Material \cite{SM} for details)
\begin{equation}
\label{T-matrix}
\mathsf{T}(\omega)
= \begin{pmatrix}
    \mathsf{T}_{++}(\omega) & \mathsf{T}_{+-}(\omega) \\
    \mathsf{T}_{-+}(\omega) & \mathsf{T}_{--}(\omega)
  \end{pmatrix}
= \overset{\leftarrow}{\cX}
	\ee^{\ii \int_{x_1}^{x_2} \dd s\, \mathsf{P}_{\omega}(s)},
\end{equation}
using the spatial ordering introduced above.
Here, manifest properties of $\mathsf{P}_{\omega}(x)$ in \eqref{P_omega_x} imply that
$\det{\mathsf{T}(\omega)} = 1$,
$\overline{\mathsf{T}(\omega)} = \mathsf{T}(-\omega)$, and
$\mathsf{T}(\omega)^{\dagger} = \sigma_{3} \mathsf{T}(\omega)^{-1} \sigma_{3}$.

The scattering matrix $\mathsf{S}(\omega)$ is obtained from $\mathsf{T}(\omega)$ by viewing
$( \hat{j}_{+}(x_1, \omega), \hat{j}_{-}(x_2, \omega) )^{T}$
as incident and
$( \hat{j}_{+}(x_2, \omega), \hat{j}_{-}(x_1, \omega) )^{T}$
as scattered currents (see the Supplemental Material \cite{SM} for details):
\begin{equation}
\label{S-matrix}
\mathsf{S}(\omega)
= \begin{pmatrix}
    T(\omega) & R(\omega) \\
    \tilde{R}(\omega) & T(\omega)
  \end{pmatrix},
\quad
\tilde{R}(\omega)
= - \overline{R(\omega)}\hspace{0.2mm} \frac{T(\omega)}{\,\overline{T(\omega)}\,}
\end{equation}
with the transmission and reflection amplitudes $T(\omega) = \frac{1}{\mathsf{T}_{--}(\omega)}$ and $R(\omega) = \frac{\mathsf{T}_{+-}(\omega)}{\mathsf{T}_{--}(\omega)}$, cf.\ \cite{Redh:1961}.
Unitarity of $\mathsf{S}(\omega)$ and $|T(\omega)|^2 + |R(\omega)|^2 = 1$ are manifest due to properties of $\mathsf{T}(\omega)$.

In principle, although nontrivially, these matrices can be computed using \eqref{int_pic_Magnus_exp}--\eqref{int_pic_Magnus_coeffs} for arbitrary $\omega$.
An important simplification is $\omega = 0$, for which $\mathsf{T}(0) = \mathsf{T}(x_2, x_1)$ in \eqref{T-matrix_xy} and
\begin{equation}
\label{T0_R0}
T(0) = \frac{2\sqrt{K(x_1)K(x_2)}}{K(x_1) + K(x_2)},
\;\;
R(0) = \frac{K(x_1) - K(x_2)}{K(x_1) + K(x_2)}
\end{equation}
for all $K(x)$.
The latter are real and depend solely on the endpoints, as the only nonvanishing zero-frequency elements in the exponential in \eqref{T-matrix} are integrals of $\Lambda(x) = \partial_x \log \sqrt{K(x)}$, yielding a simple proof of the independence on $K(x)$ and $v(x)$ for $x \in (x_1, x_2)$.
Thus, $T(0) = 1$ and $R(0) = 0$ if $K(x_2) = K(x_1)$.
Complementary arguments were given already in \cite{MaSt:1995, SaSc:1995, Safi:1997, Safi:1999}, but our results directly show that the nontrivial appearance of $K(x)$ in the transfer and scattering matrices can be attributed to integrals of $\Lambda(x)$, responsible for the gap in \eqref{iDBdG_eqs} and the new coupling in \eqref{trho_comm_rels_b}.
Moreover, for static systems, only the $\omega = 0$ contribution is nonzero, meaning that $T(0)$ and $R(0)$ give the full description of the (Andreev) scattering process.
We stress that \eqref{S-matrix} and \eqref{T0_R0} precisely reproduce the results in \cite{BachasBrunner:2008} for conformal interfaces and generalize them to any inhomogeneous compactification radius $R(x) \propto \sqrt{K(x)}$.

At $\omega = 0$, we showed that the transfer and scattering matrices for $\tilde{j}_{\pm}$ depend only on $K(x_{1, 2})$.
This has important implications for
$\rho(x, t)
= \int_{\mathbb{R}} \frac{\dd \omega}{2\pi} \hat{\rho}(x, \omega) \ee^{-\ii \omega t}$ 
and
$\jmath(x, t)
= \int_{\mathbb{R}} \frac{\dd \omega}{2\pi} \hat{\jmath}(x, \omega) \ee^{-\ii \omega t}$
in \eqref{rho_j_rho5_j5_a}.
As a corollary, their zero-frequency transfer matrix is given by
\begin{equation}
\label{T-matrix_rhoj_omega0}
\begin{pmatrix}
  \langle \hat{\rho}(x_2, 0) \rangle \\
  \langle \hat{\jmath}(x_2, 0) \rangle
\end{pmatrix}
= \begin{pmatrix}
    \frac{K(x_2)/v(x_2)}{K(x_1)/v(x_1)} & 0 \\
    0 & 1
  \end{pmatrix}
  \begin{pmatrix}
    \langle \hat{\rho}(x_1, 0) \rangle \\
    \langle \hat{\jmath}(x_1, 0) \rangle
  \end{pmatrix}.
\end{equation}
Thus, a static density profile for general $K(x)$ and $v(x)$ is affected by the latter, while the associated static current is not.
This shows the universality of the expectations $\langle \jmath_{5} \rangle = \frac{v(x)}{K(x)} \langle \rho \rangle$ and $\langle \jmath \rangle$ for arbitrary steady states since they are independent of the inhomogeneities \cite{MaSt:1995, SaSc:1995, Pono:1995, Safi:1997}.

\emph{Interacting quantum Hall edges.}---%
\csname phantomsection\endcsname%
\addcontentsline{toc}{section}{Interacting quantum Hall edges}%
%
As an application of our results, we consider edge excitations of FQH systems, see Fig.~\ref{Fig:Illustrations}(b).
We effectively describe them as chiral 1+1-dimensional CFTs of anyons \cite{Wen:1990, Wen:1992}:
For simplicity, consider Abelian anyons with statistics parameter equal to $\nu > 0$, the filling fraction of the FQH state (see, e.g., \cite{MYKGF:1993, LBFS:2009}), propagating in opposite directions along two adjacent edges,
\begin{equation}
H_{\pm}
= \int_{S^1} \dd x\,
		\anyonWick{
			\psi_{\pm}(x)^{\dagger} (\mp \ii v_{0} \partial_{x}) \psi_{\pm}(x)
		}{}.
\end{equation}
Here, $v_{0}$ is the bare velocity and $N[\cdots]$ indicates anyon normal ordering, where the latter allows one to write the associated densities as $\rho_{\pm}(x) = \anyonWick{ \psi_{\pm}(x)^{\dagger} \psi_{\pm}(x) }{}$ \cite{FreMoo:2021}.
The fields carry fractional charge $\nu e_{0}$ ($e_{0}$ denotes the elementary charge) \cite{Note:Charge_operator} and satisfy
\begin{subequations}
\begin{align}
\psi_{\pm}(x) \psi_{\pm}(y)
& = \ee^{\mp \ii \pi \nu \sgn(x-y) }
		\psi_{\pm}(y) \psi_{\pm}(x), \\
\psi_{\pm}(x) \psi_{\mp}(y)
& = \ee^{\mp \ii \pi \nu}
		\psi_{\mp}(y) \psi_{\pm}(x)
\end{align}
\end{subequations}
for $x \neq y$, with opposite signs in the exponentials when replacing $\psi_{\pm}(x)$ by $\psi_{\pm}(x)^{\dagger}$.
The edges are coupled via a density-density (four-anyon) interaction,
\begin{equation}
\label{H_anyons}
H
= H_{+} + H_{-}
  + 2\pi v_{0} \int_{S^1} \dd x\, \lambda(x) \rho_{+}(x) \rho_{-}(x)
\end{equation}
with an inhomogeneous $\lambda(x)$ satisfying $|\lambda(x)| < 1$.
In the experiment in \cite{GulEtAl:2020}, the interaction between the edges is via a superconductor, which corresponds to a Josephson coupling in the Hamiltonian, opening up a gap.
However, even without this coupling, our results imply that there is another mechanism that, if an inhomogeneous interaction $\lambda(x)$ can be realized, opens up a local gap in the governing DBdG equations, which also leads to Andreev reflections between FQH edges.

Indeed, using the boson-anyon correspondence \cite{Iso:1995, CareyLangmann:1999, FreMoo:2021}, \eqref{H_anyons} can be mapped precisely to \eqref{H_iTLL} with
\begin{equation}
\label{anyon_vx_Kx}
v(x)
= v_{0} \sqrt{ 1 - \lambda(x)^2 },
\quad
K(x)
= \sqrt{ \frac{1 - \lambda(x)}{1 + \lambda(x)} },
\end{equation}
where
$\partial_{x} \varphi(x) = - \pi[ \rho_{+}(x) - \rho_{-}(x) ]$
and
$\Pi(x) = \rho_{+}(x) + \rho_{-}(x)$.
Consider the idealized setup of two quantum Hall edges interacting as in \eqref{H_anyons} to the left,
$\lambda(x < 0^{-}) = \lambda \neq 0$,
but not to the right,
$\lambda(x > 0^{+}) = 0$.
(This captures that $\lambda(x)$ decays exponentially with distance between the edges, cf.\ \cite{GulEtAl:2020}.)
Then \eqref{T0_R0} implies
$T(0)
= \frac{2 \sqrt[4]{1 - \lambda^2}}{\sqrt{1 - \lambda} + \sqrt{1 + \lambda}}$
and
$R(0)
= \frac{\sqrt{1 - \lambda} - \sqrt{1 + \lambda}}{\sqrt{1 - \lambda} + \sqrt{1 + \lambda}}$ for $x_2 > 0 > x_1$, showing the presence of Andreev reflections since $R(0) \neq 0$.

\emph{Charge transport in quantum wires.}---%
\csname phantomsection\endcsname%
\addcontentsline{toc}{section}{Charge transport in quantum wires}%
%
Lastly, we show that our results reproduce known static results for quantum wires and allow for generalizations to dynamics following quantum quenches.
To this end, consider a quantum wire coupled to leads, the left with constant chemical potential $\mu_{L}$, velocity $v_{L}$, and Luttinger parameter $K_{L}$, and the right with $\mu_{R}$, $v_{R}$, and $K_{R}$.
This can be modeled as an inhomogeneous TLL with two well-separated steplike changes in $K(x)$, the wire identified as the part between, extending \cite{MaSt:1995, SaSc:1995, Pono:1995, GGM:2010} to general $K(x)$, see Fig.~\ref{Fig:Illustrations}(a).
One directly infers from \eqref{T-matrix_rhoj_omega0} that the static current flowing through the system is $\frac{\mu_{+} - \mu_{-}}{2\pi}$ with effective chemical potentials $\mu_{+} = K_{L} \mu_{L}$ and $\mu_{-} = K_{R} \mu_{R}$ for right movers coming from the left and vice versa, leading to the universal electrical conductance $G = \frac{e_{0}^2}{2\pi \hbar}$ for electrons (dimensionful quantities inserted) \cite{TaruchaEtAl:1995, MaSt:1995, SaSc:1995, Pono:1995, AlekseevEtAl:1996, Kawabata:1996, LLMM1:2017, GLM:2018, Note:Anyon_conductance}.

The universality of $G$ can also be obtained dynamically:
Consider a quantum quench turning off a smooth chemical-potential profile $\mu(x)$ at $t = 0$ that, outside a finite interval around $x = 0$, equals $\mu_{L}$ ($\mu_{R}$) to the left (right) \cite{LLMM1:2017, GLM:2018}.
Also, suppose $K(x)$, $v(x)$ equal $K_{L}$, $v_{L}$ ($K_{R}$, $v_{R}$) to the left (right).
Since $\frac{v(x)}{K(x)} \langle \rho \rangle$ is universal [see \eqref{T-matrix_rhoj_omega0}] and the system for $t < 0$ is in equilibrium,
$\langle \rho(y, 0) \rangle = \frac{K(y)}{\pi v(y)} \mu(y)$
and
$\langle \jmath(y, 0) \rangle = 0$,
which inserted into \eqref{j_xt_late_times} yields
$\lim_{t \to \infty} \langle \jmath(x, t) \rangle = \frac{\mu_{+} - \mu_{-}}{2\pi}$
for $|x| < \infty$.

The above is directly generalizable to inhomogeneous interaction quenches (cf.\ \cite{DoraPollmann:2015}), changing $K_{1}(x)$ to $K_{2}(x)$ at $t = 0$, corresponding to an inhomogeneous marginal ($J\bar{J}$) deformation \cite{Note:Non-exactly-marginal}, cf.\ \cite{DLMT:2023}.
For completeness, $v_{1}(x)$ is also changed to $v_{2}(x)$.
As an example, from \eqref{j_xt_late_times} with $K_{2}(x)$ and $v_{2}(x)$ and using
$\langle \rho(y, 0) \rangle = \frac{K_{1}(y)}{\pi v_{1}(y)} \mu(y)$
and
$\langle \jmath(y, 0) \rangle = 0$,
we obtain
$\lim_{t \to \infty} \langle \jmath(x, t) \rangle = \frac{\mu_{+} - \mu_{-}}{2\pi}$
for $|x| < \infty$ with
$\mu_{\pm} = \frac{K_{1}(\mp\infty) v_{2}(\mp\infty)}{v_{1}(\mp\infty)} \mu(\mp\infty)$,
similar to the nonequilibrium steady current following a homogeneous interaction quench \cite{LLMM1:2017}.

\emph{Conclusions.}---%
\csname phantomsection\endcsname%
\addcontentsline{toc}{section}{Conclusions}%
%
We developed an analytical approach to inhomogeneous TLLs (compactified free bosons) with general $v(x)$ and $K(x)$ by mapping the dynamics to inhomogeneous DBdG equations with an effective local gap and solving them exactly.
The main results are the exact Green's functions and scattering matrix, describing the nonequilibrium dynamics and showing the presence of Andreev reflections.
These generalize earlier results for conformal interfaces in boundary CFT and universal conductance in quantum wires to general inhomogeneous Luttinger parameter $K(x)$ [compactification radius $R(x) \propto \sqrt{K(x)}$] and relate them to the opening of a local gap in the governing DBdG equations.
As an application, we used our approach to study an anyonic CFT with inhomogeneous interactions as a toy model for coupled FQH edges.

One advantage of our solution is its simple description of the late-time dynamics, e.g., following a quantum quench.
These results are fully explicit and exhibit a remarkably universal dependence on $v(x)$ and $K(x)$.
We also expect our solution to be useful whenever \mbox{(Dirac-)}Bogoliubov-de~Gennes-type equations appear.
This includes the equations of motion for Majorana-Weyl fermions in \cite{GHC:2017}, which can be shown to lead to \eqref{iDBdG_eqs_omega} with $\mathsf{P}_{\omega}(x)$ generalized to lie in $\gl(2,\mathbb{C})$, and it would be interesting to apply our approach for this and other Lie algebras.
It would also be interesting to extend to Floquet drives modulating $K(x)$ in time \cite{BukovHeyl:2012dll, CitroEtAl:2020, FazziniEtAl:2021dll, DLMT:2023}, generalizing \cite{WeWu2:2018, LCTTNC1:2020, FGVW:2020, LapMoo:2021} for
time modulations of $v(x)$.
Other future directions include the disordered case of $K(x)$ given by a random function, extending \cite{LaMo2:2019}, dirty TLLs or quantum wires with noisy leads \cite{ApelRice:1982, OgataFukuyama:1994, Maslov:1995, NgoDinh:2010, StrkaljEtAl:2019, KhedriEtAl:2021}, and multicomponent TLLs, cf., e.g., \cite{Sukhorukov:2016, RCDD:2020}.

\paragraph*{Acknowledgments.}

\begin{acknowledgments}
I am thankful to Eddy Ardonne, Johan Carlstr\"{o}m, Shouvik Datta, J\'{e}r\^{o}me Dubail, Ruihua Fan, Luca Fresta, Oleksandr Gamayun, Marek Gluza, Gian Michele Graf, Andisheh Khedri, Edwin Langmann, Paola Ruggiero, Spyros Sotiriadis, and Jean-Marie St\'{e}phan for helpful and inspiring discussions.
Financial support from the Wenner-Gren Foundations (Grant No.\ WGF2019-0061) is gratefully acknowledged.
\end{acknowledgments}



\onecolumngrid
\clearpage


\section{Supplemental Material}


Part~A contains computational details for the derivation of the Green's functions in Eqs.~\eqref{j_pm_solutions}--\eqref{G_pm_xy_omega} and the transfer and scattering matrices in Eqs.~\eqref{T-matrix} and \eqref{S-matrix}, while Part~B analyzes the late-time dynamics, which, among others, leads to the late-time formula for the current in Eq.~\eqref{j_xt_late_times}.


\bigskip

\begin{center}
\bfseries\small Part~A: Green's functions and transfer and scattering matrices
\end{center}

\smallskip


Given $\mathsf{P}_{\omega}(x)$ in \eqref{P_omega_x}, define
\begin{equation}
\Xi_{\omega}(x)
\equiv
\overset{\rightarrow}{\cX} \ee^{- \ii \int_{-\infty}^{x} \dd s\, \mathsf{P}_{\omega}(s)},
\qquad
\Xi_{\omega}(x)^{-1}
= 
\overset{\leftarrow}{\cX} \ee^{\ii \int_{-\infty}^{x} \dd s\, \mathsf{P}_{\omega}(s)},
\end{equation}
where, as in the main text, $\overset{\leftarrow}{\cX}$ ($\overset{\rightarrow}{\cX}$) denotes spatial ordering where positions decrease (increase) from left to right.
It follows that \eqref{iDBdG_eqs_omega} can be written as
\begin{equation}
\partial_{x}
\biggl[
\Xi_{\omega}(x)
\begin{pmatrix}
  \langle \hat{j}_{+}(x, \omega) \rangle \\
  \langle \hat{j}_{-}(x, \omega) \rangle
\end{pmatrix}
\biggr]
= \Xi_{\omega}(x) \frac{1}{v(x)} \sigma_{3}
  \begin{pmatrix}
    \langle \tilde{j}_{+}(x, 0) \rangle \\
    \langle \tilde{j}_{-}(x, 0) \rangle
  \end{pmatrix}.
\end{equation}
Integrating from some arbitrary $a \in \mathbb{R}$ to $x$ yields
\begin{equation}
\label{DBdG_gen_solution}
\begin{pmatrix}
  \langle \hat{j}_{+}(x, \omega) \rangle \\
  \langle \hat{j}_{-}(x, \omega) \rangle
\end{pmatrix}
= \Xi_{\omega}(x)^{-1} \Xi_{\omega}(a)
  \begin{pmatrix}
    \langle \hat{j}_{+}(a, \omega) \rangle \\
    \langle \hat{j}_{-}(a, \omega) \rangle
  \end{pmatrix}
  +
  \int_{a}^{x} \dd y\,
  \Xi_{\omega}(x)^{-1} \Xi_{\omega}(y)
  \frac{1}{v(y)} \sigma_{3}
  \begin{pmatrix}
    \langle \tilde{j}_{+}(y, 0) \rangle \\
    \langle \tilde{j}_{-}(y, 0) \rangle
  \end{pmatrix},
\end{equation}
where if $a < x$, then $y < x$ in the second term and 
\begin{subequations}
\label{Xi_cases}
\begin{align}
\Xi_{\omega}(x)^{-1} \Xi_{\omega}(y)
& = \overset{\leftarrow}{\cX} \ee^{\ii \int_{y}^{x} \dd s\, \mathsf{P}_{\omega}(s)}, 
    \label{Xi_cases_1} \\
\intertext{while if $ a > x$, then $y > x$ in the second term and}
\Xi_{\omega}(x)^{-1} \Xi_{\omega}(y)
& = \overset{\rightarrow}{\cX} \ee^{-\ii \int_{x}^{y} \dd s\, \mathsf{P}_{\omega}(s)}
  = \overset{\rightarrow}{\cX} \ee^{\ii \int_{y}^{x} \dd s\, \mathsf{P}_{\omega}(s)}.
    \label{Xi_cases_2}
\end{align}
\end{subequations}
We consider the following two cases:

(i) Suppose $\langle \tilde{j}_{\pm}(x, 0) \rangle$ have compact support, consistent with assuming $\langle \hat{j}_{\pm}(\infty, \omega) \rangle = 0 = \langle \hat{j}_{\pm}(-\infty, \omega) \rangle$,  and set first $a = -\infty$ and second $a = \infty$ in \eqref{DBdG_gen_solution}.
Then we obtain
\begin{subequations}
\begin{align}
\begin{pmatrix}
  \langle \hat{j}_{+}(x, \omega) \rangle \\
  \langle \hat{j}_{-}(x, \omega) \rangle
\end{pmatrix}
& = \int_{-\infty}^{x} \dd y\,
    \overset{\leftarrow}{\cX} \ee^{\ii \int_{y}^{x} \dd s\, \mathsf{P}_{\omega}(s)}
    \frac{1}{v(y)} \sigma_{3}
    \begin{pmatrix}
      \langle \tilde{j}_{+}(y, 0) \rangle \\
      \langle \tilde{j}_{-}(y, 0) \rangle
    \end{pmatrix} \\
\intertext{and}
\begin{pmatrix}
  \langle \hat{j}_{+}(x, \omega) \rangle \\
  \langle \hat{j}_{-}(x, \omega) \rangle
\end{pmatrix}
& = -
    \int_{x}^{\infty} \dd y\,
    \overset{\rightarrow}{\cX} \ee^{\ii \int_{y}^{x} \dd s\, \mathsf{P}_{\omega}(s)}
    \frac{1}{v(y)} \sigma_{3}
    \begin{pmatrix}
      \langle \tilde{j}_{+}(y, 0) \rangle \\
      \langle \tilde{j}_{-}(y, 0) \rangle
    \end{pmatrix},
\end{align}
\end{subequations}
where we used \eqref{Xi_cases_1} and \eqref{Xi_cases_2}, respectively.
As discussed in the main text, the first solution is spatially retarded and the second is spatially advanced, and combining them so that the full result is causal yields \eqref{j_pm_solutions}--\eqref{G_pm_xy_omega}.
(Simply adding the two solutions would yield a result that contains contributions that propagate both forward and backward in time, while here we are only interested in the solution propagating forward for $t > 0$.)

(ii) Suppose $\langle \tilde{j}_{\pm}(y, 0) \rangle = 0$ for $y \in [x_1, x_2]$ and set $x = x_2$ and $a = x_1$ in \eqref{DBdG_gen_solution}.
Then we obtain
\begin{equation}
\begin{pmatrix}
  \langle \hat{j}_{+}(x_2, \omega) \rangle \\
  \langle \hat{j}_{-}(x_2, \omega) \rangle
\end{pmatrix}
= \overset{\leftarrow}{\cX} \ee^{\ii \int_{x_1}^{x_2} \dd s\, \mathsf{P}_{\omega}(s)}
  \begin{pmatrix}
    \langle \hat{j}_{+}(x_1, \omega) \rangle \\
    \langle \hat{j}_{-}(x_1, \omega) \rangle
  \end{pmatrix},
\end{equation}
where we used \eqref{Xi_cases_1}.
This yields \eqref{T-matrix} by identifying $\overset{\leftarrow}{\cX} \ee^{\ii \int_{x_1}^{x_2} \dd s\, \mathsf{P}_{\omega}(s)}$ as the transfer matrix $\mathsf{T}(\omega)$.
Reshuffling the above into
\begin{equation}
\begin{pmatrix}
  \langle \hat{j}_{+}(x_2, \omega) \rangle \\
  \langle \hat{j}_{-}(x_1, \omega) \rangle
\end{pmatrix}
= \begin{pmatrix}
    \frac{\mathsf{T}_{++}(\omega)\mathsf{T}_{--}(\omega) - \mathsf{T}_{+-}(\omega)\mathsf{T}_{-+}(\omega)}{\mathsf{T}_{--}(\omega)} & \frac{\mathsf{T}_{+-}(\omega)}{\mathsf{T}_{--}(\omega)} \\
    - \frac{\mathsf{T}_{-+}(\omega)}{\mathsf{T}_{--}(\omega)} & \frac{1}{\mathsf{T}_{--}(\omega)}
  \end{pmatrix}
  \begin{pmatrix}
    \langle \hat{j}_{+}(x_1, \omega) \rangle \\
    \langle \hat{j}_{-}(x_2, \omega) \rangle
  \end{pmatrix}
\end{equation}
yields \eqref{S-matrix} by correspondingly identifying the scattering matrix $\mathsf{S}(\omega)$ and using that the properties for $\mathsf{T}(\omega)$ stated in the main text imply that $\mathsf{T}_{++}(\omega)\mathsf{T}_{--}(\omega) - \mathsf{T}_{+-}(\omega)\mathsf{T}_{-+}(\omega) = 1$ and $\mathsf{T}_{-+}(\omega) = \overline{\mathsf{T}_{+-}(\omega)}$.


\bigskip

\begin{center}
\bfseries\small Part~B: Late-time dynamics
\end{center}

\smallskip


Recalling that the spatially retarded/advanced Green's functions are given by
$\hat{G}_{\pm}^{0}(x, y; \omega)
= \pm \theta(\pm[x-y]) \ee^{\ii \omega \tau_{x, y} \sigma_{3}} \sigma_{3}$
with $\tau_{x, y} = \int_{y}^{x} \dd s\, \frac{1}{v(s)}$
when $K(x)$ is constant, it follows from \eqref{G_pm_xy_omega} and \eqref{int_pic_Magnus_exp} that
\begin{equation}
\hat{G}_{\pm}(x, y; \omega)
= \pm \theta(\pm[x-y])
  \exp
  \biggl[ \,
    \sum_{n = 1}^{\infty}
    \Omega_{\omega}^{(n)}(x, y; x)
  \biggr]
  \ee^{\ii \omega \tau_{x, y} \sigma_{3}}
  \sigma_{3}
= \exp
  \biggl[ \,
    \sum_{n = 1}^{\infty}
    \Omega_{\omega}^{(n)}(x, y; x)
  \biggr]
  \hat{G}_{\pm}^{0}(x, y; \omega).
\end{equation}
Since, on general grounds, the late-time asymptotics is expected to be governed by the Fourier transform at small frequencies, we are motivated to consider
\begin{align}
\label{I_pm_xy_t}
I_{\pm}(x, y; t)
& \equiv
  \int_{-\infty}^{\infty} \frac{\dd \omega}{2\pi}\,
  \biggl(
  \hat{G}_{\pm}(x, y; \omega) - \ee^{\Omega_{0}^{(1)}(x, y; x)} \hat{G}_{\pm}^{0}(x, y; \omega)
  \biggr)
  \ee^{-\ii \omega t} \nonumber \\
& =
  \int_{-\infty}^{\infty} \frac{\dd \omega}{2\pi}\,
  \biggl( \exp
    \biggl[ \,
      \sum_{n = 1}^{\infty}
      \Omega_{\omega}^{(n)}(x, y; x)
    \biggr]
    - \mathsf{T}(x, y)
  \biggr)
  \hat{G}_{\pm}^{0}(x, y; \omega) \ee^{-\ii \omega t},
\end{align}
where we used that $\ee^{\Omega_{0}^{(1)}(x, y; x)} = \mathsf{T}(x, y)$ given by \eqref{T-matrix_xy}.
We seek to show that $I_{\pm}(x, y; t) \to 0$ as $t \to \infty$.
To this end, rescale $t$ by replacing it by $\lambda t$ for some $\lambda > 0$ and change the integration variable from $\omega$ to $\omega' = \lambda \omega$:
\begin{equation}
\label{I_pm_xy_lambda_t}
I_{\pm}(x, y; \lambda t)
=
\int_{-\infty}^{\infty} \frac{\dd \omega'}{2\pi\lambda}\,
\biggl(
  \exp
  \biggl[ \,
    \sum_{n = 1}^{\infty}
    \Omega_{\omega'/\lambda}^{(n)}(x, y; x)
  \biggr]
  - \mathsf{T}(x, y)
\biggr)
\hat{G}_{\pm}^{0}(x, y; \omega'/\lambda) \ee^{-\ii \omega' t}.
\end{equation}
Expanding in $\omega'/\lambda$, we have
\begin{align}
& \biggl(
    \exp
    \biggl[ \,
      \sum_{n = 1}^{\infty}
      \Omega_{\omega'/\lambda}^{(n)}(x, y; x)
    \biggr]
    - \mathsf{T}(x, y)
  \biggr)
  \hat{G}_{\pm}^{0}(x, y; \omega'/\lambda) \nonumber \\
& \qquad \qquad \qquad
  = \frac{\omega'}{\lambda}
    \partial_{\omega}
    \biggl(
      \exp
      \biggl[ \,
        \sum_{n = 1}^{\infty}
        \Omega_{\omega}^{(n)}(x, y; x)
      \biggr]
      - \mathsf{T}(x, y)
    \biggr)
    \hat{G}_{\pm}^{0}(x, y; \omega) \bigg|_{\omega = 0}
    + O(\lambda^{-2}) \nonumber \\
& \qquad \qquad \qquad
  = \pm \frac{\omega'}{\lambda} \theta(\pm[x-y]) \partial_{\omega}
    \biggl[ \,
      \sum_{n = 1}^{\infty}
      \Omega_{\omega}^{(n)}(x, y; x)
    \biggr] \bigg|_{\omega = 0}
    \mathsf{T}(x, y) \sigma_{3}
    + O(\lambda^{-2})
\end{align}
with $\Omega_{\omega}^{(n)}(x, y; x)$ given by \eqref{int_pic_Magnus_coeffs}.
Inserted into \eqref{I_pm_xy_lambda_t}, this formally implies
\begin{equation}
I_{\pm}(x, y; \lambda t)
= o(\lambda^{-1}),
\end{equation}
from which the statement on the late-time asymptotics in the main text follows.
The result for the current $\jmath$ in \eqref{j_xt_late_times} can be obtained by combing this statement with the expression $\jmath = \sqrt{K(x)} \bigl( \tilde{j}_{+} + \tilde{j}_{-} \bigr)$ in terms of $\tilde{j}_{\pm}$.

As mentioned in the main text, the above can also be verified by instead carrying out a Magnus expansion with $\omega$ as the small parameter.
Those results will be presented elsewhere.



\begin{thebibliography}{}
\bibitem{Note:Propagation_classification}
From $\sl(2, \mathbb{C})$ considerations, the propagation can be classified by $D \equiv \det[ \ii \mathsf{P}_{\omega}(x) ] = \omega^2/v(x)^2 - \Lambda(x)^2$, showing that it can be hyperbolic ($D < 0$) besides elliptic ($> 0$) or parabolic ($= 0$) [the only possibilities for constant $K(x)$].
\bibitem{Note:Time_rev_inv}
The expansion looks the same for both orderings and satisfies
$\overset{\rightarrow}{\cX} \ee^{\ii \int_{x}^{y} \dd s\, \mathsf{P}_{\omega}(s)}
= \bigl( \overset{\leftarrow}{\cX} \ee^{\ii \int_{y}^{x} \dd s\, \mathsf{P}_{\omega}(s)} \bigr)^{-1}$ since time reversibility implies
$\ee^{\ii \omega \tau_{x, y} \sigma_{3}}
\Omega_{\omega}^{(n)}(y, x; y)
\ee^{-\ii \omega \tau_{x, y} \sigma_{3}}
+ \Omega_{\omega}^{(n)}(x, y; x) = 0$,
which is manifest in \eqref{int_pic_Magnus_coeffs}.
\bibitem{Note:Anyon_conductance}
For (single-component) Abelian anyons, the conductance is $\frac{\nu e_{0}^2}{2\pi \hbar}$ since they carry charge $\nu e_{0}$.
\end{thebibliography}
\end{document}